

\input harvmac

\noblackbox
\baselineskip 20pt plus 2pt minus 2pt

\overfullrule=0pt



\def\bs{\bigskip}
\def\no{\noindent}
\def\hb{\hfill\break}
\def\qq{\qquad}
\def\bl{\bigl}
\def\br{\bigr}

\def\IR{\relax{\rm I\kern-.18em R}}

\def\np {  Nucl. Phys. }


\def\r{\rho}
\def\a{\alpha}
\def\A{\Alpha}
\def\b{\beta}

\def\d{\delta}

\def\e{\epsilon}

\def\th{\theta}

\def\m{\mu}
\def\n{\nu}

\def\l{\lambda}
\def\L{\Lambda}
\def\s{\sigma}

\def\IR{\relax{\rm I\kern-.18em R}}

\def \bd {\bar \del}
\def \bh { {\bar h} }
\def \bphi { {\bar \phi} }
\def \z { {\bar z} }
\def \A { {\bar A} }
\def \J {{\bar J} }
\def \ha {{1\over 2}}

\def \ov {\over}


\lref\BSthree{I. Bars and K. Sfetsos, Mod. Phys. Lett. {\bf A7} (1992) 1091.}

\lref\BShet{I. Bars and K. Sfetsos, Phys. Lett. {\bf 277B} (1992) 269 ;
 Phys. Rev. {\bf D46} (1992) 4495.}

\lref\BSexa{I. Bars and K. Sfetsos, Phys. Rev. {\bf D46} (1992) 4510.}

\lref\SFET{K. Sfetsos, Nucl. Phys. {\bf B389} (1993) 424.}

\lref\BSslsu{I. Bars and K. Sfetsos, Phys. Lett. {\bf 301B} (1993) 183.}

\lref\BSeaction{I. Bars and K. Sfetsos, Phys. Rev. {\bf D48} (1993) 844.}

\lref\BN{ I. Bars and D. Nemeschansky, Nucl. Phys. {\bf B348} (1991) 89.}

\lref\WIT{E. Witten, Phys. Rev. {\bf D44} (1991) 314.}

 \lref\IBhet{ I. Bars, Nucl. Phys. {\bf B334} (1990) 125. }

 \lref\IBCS{ I. Bars, {\it Curved Space-time Strings and Black Holes},
in Proc.
 {\it XX$^{th}$ Int. Conf. on Diff. Geometrical Methods in Physics}, eds. S.
 Catto and A. Rocha, Vol. 2, p. 695, (World Scientific, 1992).}

 \lref\CRE{M. Crescimanno, Mod. Phys. Lett. {\bf A7} (1992) 489.}

\lref\clapara{K. Bardakci, M. Crescimanno and E. Rabinovici,
Nucl. Phys. {\bf B344} (1990) 344.}

\lref\MSW{G. Mandal, A. Sengupta and S. Wadia,
Mod. Phys. Lett. {\bf A6} (1991) 1685.}

 \lref\HOHO{J. B. Horne and G. T. Horowitz, Nucl. Phys. {\bf B368} (1992) 444.}

 \lref\FRA{E. S. Fradkin and V. Ya. Linetsky, Phys. Lett. {\bf 277B}
          (1992) 73.}

 \lref\ISH{N. Ishibashi, M. Li, and A. R. Steif,
         Phys. Rev. Lett. {\bf 67} (1991) 3336.}

 \lref\HOR{P. Horava, Phys. Lett. {\bf 278B} (1992) 101.}

 \lref\RAI{E. Raiten, ``Perturbations of a Stringy Black Hole'',
         Fermilab-Pub 91-338-T.}

 \lref\GER{D. Gershon, ``Exact Solutions of Four-Dimensional Black Holes in
         String Theory'', TAUP-1937-91.}

 \lref \GIN {P. Ginsparg and F. Quevedo,  Nucl. Phys. {\bf B385} (1992) 527. }

 \lref\HOHOS{ J. H. Horne, G. T. Horowitz and A. R. Steif, Phys. Rev. Lett.
 {\bf 68} (1991) 568.}

 \lref\groups{
 M. Crescimanno. Mod. Phys. Lett. {\bf A7} (1992) 489. \hb
 J. B. Horne and G.T. Horowitz, Nucl. Phys. {\bf B368} (1992) 444. \hb
 E. S. Fradkin and V. Ya. Linetsky, Phys. Lett. {\bf 277B} (1992) 73. \hb
 P. Horava, Phys. Lett. {\bf 278B} (1992) 101.\hb
 E. Raiten, ``Perturbations of a Stringy Black Hole'',
         Fermilab-Pub 91-338-T.\hb
 D. Gershon, ``Exact Solutions of Four-Dimensional Black Holes in
         String Theory'', TAUP-1937-91.}

\lref\NAWIT{C. Nappi and E. Witten, Phys. Lett. {\bf 293B} (1992) 309.}

\lref\FRATSE{E. S. Fradkin and A. A. Tseytlin,
Phys. Lett. {\bf 158B} (1985) 316.}

\lref\CALLAN{ C. G. Callan, D. Friedan, E. J. Martinec and M. Perry,
Nucl. Phys. {\bf B262} (1985) 593.}

\lref\DB{L. Dixon, J. Lykken and M. Peskin, Nucl. Phys.
{\bf B325} (1989) 325.}

\lref\IB{I. Bars, Nucl. Phys. {\bf B334} (1990) 125.}

\lref\BUSCHER{T. Buscher, Phys. Lett. {\bf 201B} (1988) 466.}

\lref\RV{M. Rocek and E. Verlinde, Nucl. Phys. {\bf B373} (1992) 630.}
\lref\GR{A. Giveon and M. Rocek, Nucl. Phys. {\bf B380} (1992) 128. }

\lref\nadual{X. C. de la Ossa and F. Quevedo, ``Duality Symmetries from
Non-abelian Isometries in String Theory'', NEIP-92-004.}

\lref\SENrev{A. Sen, ``Black Holes and Solitons in String Theory'',
TIFR-TH-92-57.}

\lref\TSEd{A. A. Tseytlin, Mod. Phys. Lett. {\bf A6} (1991) 1721.}

\lref\TSESC{A. S. Schwarz and A. A. Tseytlin, ``Dilaton shift under duality
and torsion of elliptic complex'', IMPERIAL/TP/92-93/01. }

\lref\Dualone{K. Meissner and G. Veneziano,
Phys. Lett. {\bf B267} (1991) 33. \hb
K. Meissner and G. Veneziano, Mod. Phys. Lett. {\bf A6} (1991) 3397. \hb
M. Gasperini and G. Veneziano, Phys. Lett. {\bf 277B} (1992) 256. \hb
M. Gasperini, J. Maharana and G. Veneziano, Phys. Lett. {\bf 296B} (1992) 51.}

\lref\Dualtwo{A. Sen,
Phys. Lett. {\bf B271} (1991) 295;\ ibid. {\bf B274} (1992) 34. \hb
A. Sen, Phys. Rev. Lett. {\bf 69} (1992) 1006. \hb
S. Hassan and A. Sen, Nucl. Phys. {\bf B375} (1992) 103. \hb
J. Maharana and J. H. Schwarz, Nucl. Phys. {\bf B390} (1993) 3.}

\lref\KIRd{E. Kiritsis, ``Exact Duality Symmetries in CFT and String Theory'',
LPTENS-92-29; CERN-TH-6797-93.}

\lref\GIPA{A. Giveon and A. Pasquinucci, ``On cosmological string backgrounds
with toroidal isometries'', IASSNS-HEP-92/55, August 1992.}

\lref\KASU{Y. Kazama and H. Suzuki, Nucl. Phys. {\bf B234} (1989) 232. \hb
Y. Kazama and H. Suzuki Phys. Lett. {\bf 216B} (1989) 112.}

\lref\WITanom{E. Witten, Comm. Math. Phys. {\bf 144} (1992) 189.}

\lref\WITnm{E. Witten, Nucl. Phys. {\bf B371} (1992) 191.}

\lref\IBhetero{I. Bars, Phys. Lett. {\bf 293B} (1992) 315.}

\lref\IBerice{I. Bars, {\it Superstrings on Curved Space-times}, Lecture
delivered at the Int. workshop on {\it String Quantum Gravity and Physics
at the Planck Scale}, Erice, Italy, June 1992.}

\lref\DVV{R. Dijkgraaf, E. Verlinde and H. Verlinde, Nucl. Phys. {\bf B371}
(1992) 269.}

\lref\TSEY{A. A. Tseytlin, Phys. Lett. {\bf 268B} (1991) 175.}

\lref\JJP{I. Jack, D. R. T. Jones and J. Panvel,
          Nucl. Phys. {\bf B393} (1993) 95.}

\lref\BST { I. Bars, K. Sfetsos and A. A. Tseytlin, unpublished. }

\lref\TSEYT{ A. A. Tseytlin, ``Effective Action in Gauged WZW Models
and Exact String Solutions", Imperial/TP/92-93/10.}

\lref\TSEYTt{A. A. Tseytlin, ``Conformal Sigma Models corresponding to Gauged
WZW Models'', CERN-TH.6804/93.}

 \lref\SHIF { M. A. Shifman, Nucl. Phys. {\bf B352} (1991) 87.}
\lref\SHIFM { H. Leutwyler and M. A. Shifman, Int. J. Mod. Phys. {\bf
A7} (1992) 795. }

\lref\POLWIG { A. M. Polyakov and P. B. Wiegman, Phys.
Lett. {\bf 141B} (1984) 223.  }

\lref\BCR{K. Bardakci, M. Crescimanno
and E. Rabinovici, Nucl. Phys. {\bf B344} (1990) 344. }

\lref\Wwzw{E. Witten, Commun. Math. Phys. {\bf 92} (1984) 455.}

\lref\GKO{P. Goddard, A. Kent and D. Olive, Phys. Lett. {\bf 152B} (1985) 88.}

\lref\Toda{A. N. Leznov and M. V. Saveliev, Lett. Math. Phys. {\bf 3} (1979)
489. \hb A. N. Leznov and M. V. Saveliev, Comm. Math. Phys. {\bf 74}
(1980) 111.}

\lref\GToda{J. Balog, L. Feh\'er, L. O'Raifeartaigh, P. Forg\'acs and A. Wipf,
Ann. Phys. (New York) {\bf 203} (1990) 76.; Phys. Lett. {\bf 244B}
(1990) 435.}

\lref\GWZW{ E. Witten, \np {\bf B223} (1983) 422. \hb
K. Bardakci, E. Rabinovici and B. S\"aring, Nucl. Phys. {\bf B299}
(1988) 157. \hb K. Gawedzki and A. Kupiainen, Phys. Lett. {\bf 215B}
(1988) 119 ; Nucl. Phys. {\bf B320}(1989) 625.}

\lref\SCH{ D. Karabali, Q-Han Park, H. J. Schnitzer and Z. Yang,
                   Phys. Lett. {\bf B216} (1989) 307. \hb D. Karabali
and H. J. Schnitzer, Nucl. Phys. {\bf B329} (1990) 649. }

 \lref\KIR{E. Kiritsis, Mod. Phys. Lett. {\bf A6} (1991) 2871. }

\lref\BIR{N. D. Birrell and P. C. W. Davies,
{\it Quantum Fields in Curved Space}, Cambridge University Press.}

\lref\WYB{B. G. Wybourn, {\it Classical Groups for Physicists }
(John Wiley \& sons, 1974).}

\lref\SANTA{R. Guven, Phys. Lett. {\bf 191B} (1987) 275.\hb
D. Amati and C. Klimcik, Phys. Lett. {\bf 219B} (1989) 443.\hb
G. T. Horowitz and A. R. Steif, Phys. Rev. Lett. {\bf 64} (1990) 260.}

\lref\SANT{J. H. Horne, G. T. Horowitz and A. R. Steif,
Phys. Rev. Lett. {\bf 68} (1991) 568.}

\lref\PRE{J. Prescill, P. Schwarz, A. Shapere, S. Trivedi and F. Wilczek,
Mod. Phys Lett. {\bf A6} (1991) 2353.\hb
C. Holzhey and F. Wilczek, Nucl. Phys. {\bf B380} (1992) 447.}

\lref\HAWK{J. B. Hartle and S. W. Hawking Phys. Rev. {\bf D13} (1976) 2188.\hb
S. W. Hawking, Phys. Rev. {\bf D18} (1978) 1747.}

\lref\HAWKI{S. W. Hawking, Comm. Math. Phys. {\bf 43} (1975) 199.}

\lref\HAWKII{S. W. Hawking, Phys. Rev. {\bf D14} (1976) 2460.}

\lref\euclidean{S. Elitzur, A. Forge and E. Rabinovici,
Nucl. Phys. {\bf B359} (1991) 581. }

\lref\ITZ{C. Itzykson and J. Zuber, {\it Quantum Field Theory},
McGraw Hill (1980). }

\lref\kacrev{P. Goddard and D. Olive, Journal of Mod. Phys. {\bf A} Vol. 1,
No. 2 (1986) 303.}

\lref\BBS{F. A. Bais, P. Bouwknegt and M. Surridge, Nucl. Phys. {\bf B304}
(1988) 348.}

\lref\nonl{A. Polyakov, {\it Fields, Strings and Critical Phenomena}, Proc. of
Les Houses 1988, eds. E. Brezin and J. Zinn-Justin North-Holland, 1990.\hb
Al. B. Zamolodchikov, preprint ITEP 87-89. \hb
K. Schoutens, A. Sevrin and P. van Nieuwenhuizen, Proc. of the Stony Brook
Conference {\it Strings and Symmetries 1991}, World Scientific,
Singapore, 1992. \hb
J. de Boer and J. Goeree, ``The Effective Action of $W_3$ Gravity to all
\hb orders'', THU-92/33.}

\lref\HOrev{G. T. Horowitz, {\it The Dark Side of String Theory:
Black Holes and Black Strings}, Proc. of the 1992 Trieste Spring School on
String Theory and Quantum Gravity.}

\lref\HSrev{J. Harvey and A. Strominger, {\it Quantum Aspects of Black
Holes}, Proc. of the 1992 Trieste Spring School on
String Theory and Quantum Gravity.}

\lref\GM{G. Gibbons, Nucl. Phys. {\bf B207} (1982) 337.\hb
G. Gibbons and K. Maeda, Nucl. Phys. {\bf B298} (1988) 741.}

\lref\GID{S. B. Giddings, Phys. Rev. {\bf D46} (1992) 1347.}

\lref\PRErev{J. Preskill, {\it Do Black Holes Destroy Information?},
Proc. of the International Symposium on Black Holes, Membranes, Wormholes,
and Superstrings, The Woodlands, Texas, 16-18 January, 1992.}

\lref\tye{S-W. Chung and S. H. H. Tye, Phys. Rev. {\bf D47} (1993) 4546.}

\lref\eguchi{T. Eguchi, Mod. Phys. Lett. {\bf A7} (1992) 85.}

\lref\HSBW{P. S. Howe and G. Sierra, Phys. Lett. {\bf 144B} (1984) 451.\hb
J. Bagger and E. Witten, Nucl. Phys. {\bf B222} (1983) 1.}

\lref\GSW{M. B. Green, J. H. Schwarz and E. Witten, {\it Superstring Theory},
Cambridge Univ. Press, Vols. 1 and 2, London and New York (1987).}

\lref\KAKU{M. Kaku, {\it Introduction to Superstrings}, Springer-Verlag, Berlin
and New York (1991).}

\lref\LSW{W. Lerche, A. N. Schellekens and N. P. Warner, {\it Lattices and
Strings }, Physics Reports {\bf 177}, Nos. 1 \& 2 (1989) 1, North-Holland,
Amsterdam.}

\lref\confrev{P. Ginsparg and J. L. Gardy in {\it Fields, Strings, and
Critical Phenomena}, 1988 Les Houches School, E. Brezin and J. Zinn-Justin,
eds, Elsevier Science Publ., Amsterdam (1989). \hb
J. Bagger, {\it Basic Conformal Field Theory},
Lectures given at 1988 Banff Summer Inst. on Particle and Fields,
Banff, Canada, Aug. 14-27, 1988, HUTP-89/A006, January 1989. }

\lref\CHAN{S. Chandrasekhar, {\it The Mathematical Theory of Black Holes},
Oxford University Press, 1983.}

\lref\KOULU{C. Kounnas and D. L\"ust, Phys. Lett. {\bf 289B} (1992) 56.}

\lref\PERRY{M. J. Perry and E. Teo, ``Non-singularity of the Exact two
Dimensional String Black Hole'', DAMTP-R-93-1. \hb
P. Yi, ``Nonsingular 2d Black Holes and Classical String Backgrounds'',
CALT-68-1852. }

\lref\GiKi{A. Giveon and E. Kiritsis, ``Axial Vector Duality Symmetry and
Topology Change in String Theory'', CERN-TH-6816-93.}

\lref\kar{S. K. Kar and A. Kumar, Phys. Lett. {\bf 291B} (1992) 246.}

\lref\NW{C. Nappi and E. Witten, Phys. Rev. Lett. {\bf 71} (1993) 3751.}
\lref\HK{M. B. Halpern and E. Kiritsis,
Mod. Phys. Lett. {\bf A4} (1989) 1373.}
\lref\MOR{A.Yu. Morozov, A.M. Perelomov, A.A. Rosly, M.A. Shifman
and A.V. Turbiner, Int. J. Mod. Phys. {\bf A5} (1990) 803.}
\lref\KK{E. Kiritsis and C. Kounnas, ``String propagation in Gravitational Wave
Backgrounds'',
CERN-TH.7059/93, hepth/9310202.}
\lref\KST{K. Sfetsos and A.A. Tseytlin, ``Antisymmetric tensor coupling and
conformal
invariance in sigma models corresponding to gauged WZNW theories'',
CERN-TH.6969/93,
THU-93/25, hepth/9310159, to appear in Phys. Rev. {\bf D} (1994).}

\lref\KSTh{K. Sfetsos and A.A. Tseytlin, ``Chiral gauged WZNW models and
heterotic string
backgrounds'', CERN-TH.6962/93, USC-93/HEP-S2, hepth/9308018, to appear in
Nucl. Phys. {\bf B} (1994).}

\lref\KP{S.P. Khastgir and A. Kumar, ``Singular limits and string solutions'',
IP/BBSR/93-72, hepth/9311048.}
\lref\ark{A.A. Tseytlin, Phys. Lett. {\bf B317} (1993) 559.}
\lref\GK{A. Giveon and E. Kiritsis, ``Axial Vector Duality as a gauge symmetry
and Topology change in String Theory, CERN-TH-6816-93, hepth/9303016.}

\lref\saletan{E.J. Saletan, J. Math. Phys. {\bf 2} (1961) 1.}
\lref\jao{D. Cangemi and R. Jackiw, Phys. Rev. Lett. {\bf 69} (1992) 233.}
\lref\jat{D. Cangemi and R. Jackiw, Ann. Phys. (NY) {\bf 225} (1993) 229.}
\lref\KT{C. Klimcik and A.A. Tseytlin, ``Duality invariant class of
exact string backgrounds'', CERN TH.7069, hepth/9311012.}

\lref\ORS{ D. I. Olive, E. Rabinovici and A. Schwimmer, ``A class of String
Backgrounds as a Semiclassical Limit of WZW Models'', SWA/93-94/15,
hepth/93011081.}

\lref\edc{K. Sfetsos, ``Exact String Backgrounds from WZW models based on
Non-semi-simple groups'', THU-93/31, hepth/9311093,
to appear in Int. J. Mod. Phys. {\bf A} (1994). }


\hfill {THU-93/30}
\vskip -.3 true cm
\rightline{November 1993}
\vskip -.3 true cm
\rightline {hep-th/9311010 }

\bs\bs\bs

\centerline  {\bf GAUGING A NON-SEMI-SIMPLE WZW MODEL }

\vskip 1.00 true cm

\centerline  {  {\bf Konstadinos Sfetsos}{\footnote{$^*$}
 {e-mail address: sfetsos@ruunts.fys.ruu.nl }}
   }

\bigskip

\centerline {Institute for Theoretical Physics }
\centerline {Utrecht University}
\centerline {Princetonplein 5, TA 3508}
\centerline{ The Netherlands }


\vskip 1.50 true cm

\centerline{ABSTRACT}

\vskip .3 true cm

We consider gauged WZW models based on a four dimensional
non-semi-simple group. We obtain conformal $\s$-models in $D=3$
spacetime dimensions (with exact central charge $c=3$) by axially
and vectorially gauging a one-dimensional subgroup.
The model obtained in the axial gauging is related to the $3D$ black string
after a correlated limit is taken in the latter model.
By identifying the CFT corresponding to these $\s$-models
we compute the exact expressions for the metric and dilaton fields.
All of our models can be mapped to flat
spacetimes with zero antisymmetric tensor and dilaton fields via
duality transformations.

\vfill\eject


\lref\kacm{V. G. Kac, Funct. Appl. {\bf 1} (1967) 328.\hb
           R.V. Moody, Bull. Am. Math. Soc. {\bf 73} (1967) 217.}
\lref\HB{K. Bardakci and M.B. Halpern, Phys. Rev. {\bf D3} (1971) 2493.}
\lref\MBH{M.B. Halpern, Phys. Rev {\bf D4} (1971) 2398.}
\lref\balog{J. Balog, L. O'Raifeartaigh, P. Forgacs and A. Wipf,
Nucl. Phys. {\bf B325} (1989) 225.}

\newsec{Introduction}

A large class of Conformal Field Theories (CFT's) can be constructed by using
current algebras \kacm\HB. Models where the full symmetry of the action is
realized in terms of current algebras are the WZW models \Wwzw\ based on a
group $G$. By gauging an anomaly free subgroup $H$ of $G$ we obtain new
conformal theories or coset models $G/H$ \HB\MBH\GKO.
Both WZW and gauged WZW models \GWZW\
are very important in understanding String theory since they provide exact
solutions to it and the current algebra description makes the theories
solvable.
Compact groups were used in the construction of such models in String
compactifications as internal theories where the space-time was taken to be
Minkowskian. Later non-compact groups were used for
non-compact current algebras and coset models that provided  exact String
models in {\it curved} space-time \balog\IB.
Explicit constructions of the geometries corresponding to gauged WZW models
was done in the recent years (see for example \WIT\HOHO\BShet\NAWIT)
including all order corrections to
the semiclassical (leading order) results (see for example \DVV\BSexa).

So far the attention has been concentrated to the cases where simple or
semi-simple groups were used for reasons that will become obvious.
In a recent paper Nappi and Witten \NW\ showed how to write a WZW action for a
non-semi-simple group.
They considered the algebra $E^c_2$,
i.e. the $2D$ Euclidean algebra with a central extension operator
$T$. The algebra for the generators $T_A=\{P_1,P_2,J,T\}$ is\foot{This algebra
appeared before in the context of contraction of Lie groups \saletan\ and
in studies of $(1+1)$-dimensional gravity \jao.}
\eqn\alg{[J, P_i]=\e_{ij}P_j\ ,
\qq [P_i, P_j]=\e_{ij}T \ ,\qq [T,J]=[T,P_i]=0\ .}
Unlike the case of semi-simple algebras here the quadratic form
$\Omega_{AB}=f_{AC}{}^D f_{BD}{}^C$ is degenerate, i.e. its
determinate
is zero. This makes the straightforward writing of the corresponding WZW action
problematic.
However one can still resolve this problem by considering another quadratic
form which
satisfies the properties a) $\Omega_{AB}=\Omega_{BA}$, b) $f^D_{AB}\Omega_{CD}
+ f^D_{AC}\Omega_{BD} = 0$ and c) is non-degenerate, i.e. the inverse matrix
$\Omega^{AB}$ obeying $\Omega^{AB}\Omega_{BC}=\delta^A_C$ exists.
The first and the second properties ensure the existence of the quadratic and
the Wess-Zumino
term in the WZW action and the third one gives a way to properly lower and
raise indices.
The most general such quadratic form is \jat
\eqn\quadr{\Omega_{AB}=k\pmatrix{1 & 0 & 0 & 0\cr
0 & 1 & 0 & 0\cr
0 & 0 & b & 1\cr
0 & 0 & 1 & 0\cr}\ .}
Then parametrizing the group element as (summation over repeated indices
is implied)
\eqn\group{ g= e^{a_i P_i}e^{uJ+vT} \ ,}
\no
the corresponding WZW action takes the form \NW
\eqn\action{S(u,v,a_i)={k\over 2\pi} \int d^2z \ \bl( \del a_i \bd a_i
+ \del u \bd v + \del v \bd u + b\ \del u \bd u
+ \e_{ij} \del a_i a_j \bd u \br)\ .}
\no
{}From this action one can easily read off the corresponding metric and
antisymmetric fields
which represent a monochromatic plane wave. The corresponding exact CFT was
identified as
a solution of the Master equation of the generalized virasoro construction of
\HK\MOR\ with
central charge $c=4$. The 1-loop solution is in fact exact since there are not
any higher loop
diagrams that can even be drawn \NW. In fact there is an alternative way to
understand the absence of higher loop corrections.
If we change variables as \NW
\eqn\chcoo{a_1= x_1 + x_2 \cos u\ , \qq  a_2= x_2 \cos u\ ,
\qq v\to v + \ha x_1 x_2 \sin u\ ,}
then the action \action\ reads
\eqn\act{ S={k\ov 2\pi} \int d^2z\ \bl( \del x_1 \bd x_1 + \del x_2 \bd x_2
+ 2 \cos u\ \del x_1 \bd x_2 + b\ \del b \bd u + 2 \del u \bd v \br )\ .}
In this form it can easily be shown that it is equivalent to a correlated
limit of the WZW action for $SU(2) \otimes \IR$. For the latter model, if we
parametrize the $SU(2)$ group element as
$$g= e^{i {\s_1\ov 2} \th_L}\ e^{i {\s_3\ov 2} \phi}\ e^{i {\s_1\ov 2} \th_R}$$
and the translational factor $\IR$ in terms of the time-like
coordinate $y$, the action is
\eqn\acts{S={k'\ov 4\pi} \int d^2 z\ \bl( \del \th_L \bd \th_L
+ \del \th_R \bd \th_R + \del \phi \bd \phi +2 \cos \phi\ \del\th_L \bd \th_R
- \del y \bd y \br) \ .}
If we define
$ k'= 2 k/\e$, $\th_L= \sqrt{\e}\ x_1$,
$\th_R= \sqrt{\e}\ x_2$, $\phi= \e v+ u$, $y=(1-\e b/2)\ u $
and take the limit $\e\to 0$ the action \acts\ becomes identical to \act.
The absence of higher loop corrections in \act\ is then attributed to the fact
that such corrections are also absent in the WZW action for $SU(2) \otimes \IR$
(except for a trivial overall shift in the value of $k'$). Notice also that
after the rescaling the periodic variables $\th_L$ and $\th_R$ take values,
as $x_1$ and $x_2$, in the whole real line.

As in the case of WZW models based on simple or semi-simple groups one can
obtain new conformal models by gauging anomaly free subgroups of \action.
In this paper we consider the gauging of the WZW model of \NW\
with respect to the generator $J$ of the Cartan subalgebra.
In section 2 we consider the axial, vector and chiral gauging cases.
It will be shown that our $\s$-models can also be obtained if a specific limit
in the $3D$ charged and neutral black string models based on
$SL(2,\IR)\otimes \IR/\IR$ are taken. In particular the limit is such that it
`explores' the region around the curvature singularities present
in the latter models.
By performing a duality transformation we show that all of our solutions can be
mapped to
flat spacetime solutions with zero antisymmetric tensor and constant dilaton
showing that these singularities are harmless from the point of view of String
theory.
In section 3 we are identifying the exact CFT corresponding to the $\s$-model
solution of the previous section
as a particular case of the models of \HK\MOR\ with $c=3$.
Using the Hamiltonian for this CFT we compute the exact expressions for the
metric and dilaton fields. We end this paper with some concluding remarks in
section 4.

\newsec{ Axial, vector and chiral gauging}

In this section we consider different gaugings of the WZW model for
$E^c_2$ with respect to the $U(1)$ subgroup generated by $J$,
i.e. $E^c_2/U(1)$.
A more general gauging of the linear combination $J+\l
T$ turns out to give an identical to the $\l=0$ case results, up to a shift in
the value of the constant $b$ in the expressions below (this has its origin
in the fact that the algebra \alg\ is invariant under the redefinition
$J\to J + \l T$ ).

\subsec{ Axial gauging}

Consider first the case of the axial gauging
(it turns to be the most interesting one) which is not anomalous since
the subgroup generated by $J$ is abelian.
In this case the gauged WZW action is (see for instance \WIT\GIN)
\eqn\saxial{\eqalign{
&S_{\rm axial} = k\ [\ I( h g \bh )  -  I( h\inv \bh ) \ ]\cr
& h= e^{-J \phi}\ , \qq \bh = e^{-J\bphi} \ .\cr } }
\no
The action $S_{\rm axial}$ is invariant
under the following gauge transformations
\eqn\transf{ \d a_i = - \e_{ij} a_j\ \e\ , \quad \d u = 2\ \e \ ,
\quad \d v= 0\ , \quad \d\phi =\d \bphi = \e\ ,\quad \e =\e (z,\z) \ .}
\no
The easiest way to realize that is to use the commutation relations \alg\
in order to rewrite the above action \saxial\ in the form
\eqn\actt{S_{\rm axial}=k\ [\ I(e^{a'_i P_i} e^{ u' J + v' T})
- I(e^{u'' J }) \ ] \ ,}
\no
where we have defined
\eqn\defi{
a'_i = (\cos \phi\ \d_{ij} + \sin \phi\ \e_{ij}) a_j\ , \qq
u'= u- \phi- \bphi \ , \qq v'= v\ ,\qq u''= \phi - \bphi \ .}
\no
Gauge invariance of \actt\ under \transf\ is manifest upon realizing that
$\d a'_i = 0$. Using the explicit from of the action
for the WZW model \action, and the formulae
\eqn\expl{\eqalign{&\del a'_i \bd a'_i = \del a_i \bd a_i
+ a_i a_i \del \phi \bd \phi + \e_{ij} (\del \phi \bd a_i a_j
+ \del a_i a_j \bd \phi) \cr
&\e_{ij} \del a'_i a'_j=\e_{ij} \del a_i a_j + a_i a_i \del \phi\ , \cr } }
\no
one can cast $S_{\rm axial}$ in \saxial, in the usual form of a gauged WZW
model
\eqn\gwzw{\eqalign{&S_{\rm axial}={k\over 2\pi} \int d^2z \ \bl(\
 \del a_i \bd a_i
+ \del u \bd v + \del v \bd u + b\ \del u \bd u
+ \e_{ij} \del a_i a_j \bd u  \cr
&+ A\ ( \e_{ij} \bd a_i a_j - 2 \bd v +( a_i a_i -2 b )\bd u )
-( \e_{ij} \del a_i a_j + 2 \del v + 2 b \del u )\ \A + ( 4 b  -a_i a_i) \
A\A\ \br)
\ ,\cr }}
\no
where we have defined the gauged fields as $A= \del \phi$ and
$\A = \bd \bphi $.
To obtain the $\s$-model we have to fix the gauge and integrate over the gauge
fields.
A convenient gauge choice is $a_1=0$.\foot{This gauge fixing introduces
a Faddeev-Popov factor $(FP)\sim a_2$ in the path integral measure.
This factor should combine with the original measure in the path integral for
the
gauged WZW model (in our case is the flat one) to give $e^{\Phi} \sqrt{G}$
\KIR\BSthree.
The fact that this is indeed the case can be verified using the appropriate
expressions below.}
The resulting $\s$-model action is (we will denote $\r=a_2$)
\eqn\smod{ S= {k\over 2\pi} \int d^2z\ \bl(\
 \del \r \bd \r + {4\over 4b -\r^2}
( -\del v \bd v + {b \over 4} \r^2\ \del u\bd u )
 + {4b\over 4b -\r^2} (\del v \bd u - \del u \bd v)\ \br)}
\no
Let us assume that $b>0$ and define the rescaled variables
$r={\r \ov 2 \sqrt{b}}$, $ x=v/\sqrt{b}$, $ y=\sqrt{b}\ u$.
Then the metric, the antisymmetric tensor one reads off from the action \smod\
and the dilaton induced from integrating out the gauge fields
are given by
\eqn\meax{\eqalign{&ds^2= 4b\ dr^2 +{1\over r^2-1}\ dx^2
-{r^2\over r^2-1}\ dy^2 \cr
&B_{xy}=-{1\over r^2-1}\ ,\qq B_{xr}=B_{yr}=0 \cr
&\Phi= \ln(r^2-1) +{\rm const.}\ .\cr }}
\no
Let us verify that the
above fields do in fact solve the equations for conformal invariance in the
1-loop approximation. In $D=3$ the antisymmetric field strength $H_{\m\n\r}=
3 \partial_{[\r} B_{\m\n]}$ has only one component which can be parametrized
in terms of a scalar $H$ as $H_{\m\n\r}=\e_{\m\n\r} H$.
In our case
$$H= {1\over \sqrt{G}}\ \partial_r B_{xy}=
- {i\over \sqrt{b}}\ {1\over r^2-1}\ .$$
Then in $D=3$ the 1-loop equations for conformal invariance
(see for instance \CALLAN) can be written as \KST
\eqn\bita{\eqalign{
&{\bar \b}^G_{\m\n}= R_{\m\n} -{1\over 2} H^2 G_{mn} - D_{\m} D_{\n}\Phi =0 \cr
&{\bar \b}^B_{\n\l}\e_{\m}{}^{\n\l}= - e^{-\Phi}\partial_{\m} (e^{\Phi} H ) =0
\cr
& {\bar \b}^{\Phi}={1\over 6} (3-c) + {\a'\over 4} [ D^2\Phi
+ (\del_{\m}\Phi)^2 - H^2 ] =0
\ .\cr }}
\no
We can explicitly compute the relevant tensors
\eqn\rmn{\eqalign{
&R_{rr}=-2\ {r^2+2\over (r^2-1)^2}\ ,\ R_{xx}=-{1\over2 b}\
{r^2+1\over (r^2-1)^3}\ ,\ R_{yy}={1\over b}\ {r^2\over (r^2-1)^3}\cr
&D_rD_r\Phi=-2\ {r^2+1\over (r^2-1)^2}\ ,\ D_xD_x\Phi= -{1\over 2 b}\
{r^2\over (r^2-1)^3}\ ,\ D_yD_y\Phi={1\over 2 b}\ {r^2\over (r^2-1)^3} \cr } }
\no
with the off-diagonal elements being zero and the scalars
\eqn\scal{R= -{1 \over 2 b }\ {2 r^2 +5\over (r^2-1)^2}\ ,\qq
D^2\Phi=-{1\over b}\ {r^2+1\over (r^2-1)^2} \ .}
\no
One then can easily verify that the equations \bita\ are indeed satisfied
with central charge $c=3$.
The signature of the spacetime in \meax\ is $(+,+,-)$.
If $b<0$ one has to anallytically continue $(r,x,y)\to (ir, ix, iy)$. Then the
spacetime
in \meax\ has signature $(+,+,+)$ and no singularity at all.
The background defined in \meax\ is related to the corresponding one of the
$D=3$ charged black string based on the coset model $SL(2,\IR) \otimes \IR/\IR$
through a limiting procedure. The expressions for the latter are \HOHO
\eqn\blst{\eqalign{&ds^2={1\ov \a'}(-{z-q_0-1\ov z}\ dt^2 + {z-q_0\ov z}\ dx^2
+ {dz^2 \ov 4(z-q_0-1)(z-q_0)})\cr
&B_{xt}={1\ov \a'}{\sqrt{q_0(q_0+1)}\ov z}\ , \qq
\Phi=\ln z +{\rm const.}\ .\cr } }
If we take the following correlated limits $z\to \e (r^2-1)$, $q_0\to -1-\e$,
$x\to \sqrt{\e\a'}  x$, $t\to -\sqrt{\a'} y$ with $\a'\equiv \e/{4 b}$ and
$\e\to 0$ then \blst\ becomes \meax.\foot{
It is interesting to note that there is another limiting procedure one can
follow in \blst\ (or by analytically continue \meax)
leading to a solution with a different physical interpretation.
Namely if $z\to \e(t^2+1)$, $q_0\to \e$, $x\to \sqrt{\a'} x$,
$t\to \sqrt{\a' \e} y$ with $\a'=\e/b$ and $\e\to 0$ then \blst\ becomes
$ds^2=-b dt^2+ (t^2+1)^{-1} (dy^2 + t^2 dx^2) $, $B_{xy}=1/(t^2+1)$ and
$\Phi=\ln (t^2+1)$. The above metric has the cosmological interpretation
of an expanding and recollapsing Universe.
String backgrounds obtained by considering various limiting procedures
on already existing solutions can also be found in \KP.}
One might wonder what the physical meaning of such a correlated limit is.
It is known \HOHO\ that the black string geometry \blst\ has a curvature
singularity at $z=0$. Our limit corresponds to `magnifying' the region around
$z=0$. The outer and inner horizons for $z=q_0+1$ and $z=q_0$ in the metric
\blst\ dissapear in the above limit exactly because we `look' at the region
close to $z=0$ (already inside the inner horizon).
Let us note that if we had taken the (straightforward)
limit of zero axionic charge $q_0\to -1$
(essentially zero embedding of $H=\IR$ into the $\IR$ factor in $G$ -- up to an
analytic continuation) we would have obtained the $\s$-model for the
$SL(2,\IR)/U(1) \otimes \IR$ (the Euclidean black hole times a translation).
The metric defined in \meax\ has a time-like curvature singularity at
$r^2=1$ (corresponding to the black string singularity at $z=0$).
However, from the point of view of string theory this is not very harmful
because
next we will show that the fields in \meax\ are related to the $D=3$ Minkowski
space-time (with constant dilaton and zero antisymmetric tensor) by means
of a duality transformation. In contrast the background \blst\ is dual
to the neutral black string (with $q_0=0$) which is the $2D$ black hole
times a $U(1)$. The duality transformations for the case of one
isometry read (the coordinate system is $(x^0,x^a)$) \BUSCHER
\eqn\duality{\eqalign{&\tilde G_{00}={1\over G_{00}}\ ,\quad
\tilde g_{0a}={B_{0a}\over G_{00}}\ , \quad \tilde G_{ab}=G_{ab}-
{G_{0a} G_{0b} - B_{0a} B_{0b}\over G_{00} }\cr
& \tilde B_{0a}={G_{0a}\over G_{00}}\ ,\quad \tilde B_{ab}=B_{ab}-
{G_{0a} B_{0b}-G_{0b}B_{0a}\over G_{00}}\cr
&\tilde \Phi=\Phi +\ln G_{00} \ .\cr } }
\no
Applying this transformation to the background in \meax\ for the isometry
in the $x$-direction we obtain
\eqn\res{\eqalign{
&d\tilde s^2= 4b\ dr^2 +(r^2-1)\ dx^2 -dy^2 -2\ dx dy\cr
&\tilde B_{\m\n}=0\ , \qq \tilde\Phi={\rm const.}\ .\cr} }
\no
After we make the shift of the coordinate $y\to y-x$ we see that the metric
becomes the Minkowski metric in three dimensions,
i.e. $d\tilde s^2=4b\ dr^2 +r^2 dx^2 -dy^2$.
If we apply \duality\ for the isometry along the $y$-direction we get
(we also have to shift $x\to x+y$ )
\eqn\ress{\eqalign{
&d \tilde s^2= 4b\ dr^2 +dy^2 - {1\over r^2}\ dx^2\cr
&\tilde B_{\m\n}=0\ ,\qq \tilde\Phi=\ln r^2 +{\rm const.}\ .\cr } }
\no
Obviously this solution can also be mapped into the background corresponding to
the
flat Minkowski space-time with zero antisymmetric tensor and constant dilaton
by performing a duality transformation in \ress\ along the $x$-direction.

\subsec{ Vector and chiral gaugings}

Let us now consider the case of the vector gauging. In this case the action is
\GWZW
\eqn\vector{\eqalign{&S_{\rm vector}= k\ [\ I(h\inv g \bh)
- I(h\inv \bh) \ ] \cr
& h=e^{-J\phi}\ ,\qq \bh=e^{-J\bphi}\ . \cr } }
\no
As in the case of the axial gauging it is convenient to cast \vector\ in the
form
\eqn\actv{S_{\rm vector}=k\ [\ I(e^{a'_i P_i} e^{ u' J + v' T})
- I(e^{u'' J }) \ ] \ ,}
\no
but now with definitions
\eqn\defv{a'_i = (\cos \phi\ \d_{ij} - \sin \phi\ \e_{ij}) a_j\ , \qq
u'= u + \phi- \bphi \ , \qq v'= v\ ,\qq u''= \phi - \bphi \ . }
\no
Then it is easy to check that the action \vector\ (or equivalently \actv)
is invariant under the vector-like gauge transformations
\eqn\transfv{\d a_i =  \e_{ij} a_j\ \e\ , \quad \d u=\d v =0 \ ,
\quad \d\phi =\d \bphi = \e\ ,\quad \e =\e (z,\z) \ .}
\no
Then
\eqn\svec{\eqalign{
&S_{\rm vector}={k\over 2\pi} \int d^2 z\ \bl(\ \del a_i \bd a_i
+ \del u \bd v + \del v \bd u + b\ \del u \bd u
+ \e_{ij} \del a_i a_j \bd u  \cr
&+A\ (2 \bd v +(2 b -a_i a_i)\bd u  -\e_{ij} \bd a_i a_j )
-(2 \del v +2 b \del u +\e_{ij} \del a_i a_j  ) \A +a_i a_i A\A\ \br)\ .\cr} }
\no
The result of the integration over the gauge fields is (As before we fix the
gauge by choosing $a_1=0$ and we denote $\r=a_2$. We also perform the
shifting $v\to v-bu$)
\eqn\rese{\eqalign{& ds^2= d\r^2 -b\ du^2 +{4\over \r^2} \ dv^2 \cr
& B_{\m\n}=0\ ,\qq \Phi=\ln \r^2 +{\rm const.} \ .\cr } }
\no
Perhaps this simple result should have
been expected since not only $v$ but also $u$ is inert under the vector
transformations \transfv. It represents the dual space-time to the
Euclidean space-time in $2D$ in polar coordinates \RV\ plus a time-like
coordinate. Thus, the coset $E^c_2/U(1)$ provides also a exact
CFT description for the space \rese\ as well.
In view of the fact that the background
in \meax\ can be obtained from that for the charged black string \blst\
by taken a correlated limit, one
might expect that \rese\ may be derived from the dual of the charged black
string \SANT, namely $SL(2,\IR)/\IR \otimes \IR$, via a similar limiting
procedure. One can easily verify that
(again the limit corresponds to a `maginification' of the region
around the $2D$ black hole  singularity at $uv=1$).
Notice also that \rese\ is dual to an analytically
continued version of \ress.

Let us consider briefly the case of chiral gauging which, generically,
gives rise
to $\s$-models that are inequivalent with the corresponding models one
obtains in the usual cases of axial and vector gaugings.
Nevertheless, they will also be conformally invariant because the action
one starts with, before integrating out the gauge fields,
can be written as the sum of three {\it independent} WZW actions,
as it was discussed in \KSTh.
Namely, in the chiral gauging case the action has the form \tye\
\eqn\chiral{\eqalign{
&S_{\rm chiral}=k\ [\ I(h g \bh)- I(h)-I(\bh)\ ]\cr
&={k\over 2\pi} \int d^2z \ \bl(\
 \del a_i \bd a_i
+ \del u \bd v + \del v \bd u + b\ \del u \bd u
+ \e_{ij} \del a_i a_j \bd u  \cr
&+ A\ ( \e_{ij} \bd a_i a_j - 2 \bd v +( a_i a_i -2 b )\bd u )
-( \e_{ij} \del a_i a_j + 2 \del v + 2 b \del u )\ \A
 + ( 2 b  -a_i a_i ) \ A\A\ \br) \ .\cr } }
\no
The above action is invariant under the chiral gauge transformations
\eqn\chg{\d a_i=-\e_{ij} a_j\ \e\ ,\quad \d u=\e+{\bar \e}\ ,\quad \d v=0\ ,
\quad \d \phi=\e\ ,\quad \d \bphi={\bar \e}\ , \quad \e=\e(z)\ ,
\quad {\bar \e}={\bar \e}(\z) \ .}
\no
After integrating over $A$, $\A$ and make the shift $v\to v-{b\over 2} u$ we
obtain
\eqn\cmbp{\eqalign{&ds^2= {1\over a^2 -2b}\ \bl( (a_i da_i)^2 -2b\ da_i da_i
-2b\ \e_{ij} da_i a_j du +4 dv^2 +b(b-a^2)\ du^2\ \br)\cr
&B_{uv}={a^2\over a^2-2b}\ , \qq B_{iv}={2\over a^2-2b}\ \e_{ij} a_j\cr
&\Phi=\ln(a^2-2b) +{\rm const.}\ ,\cr}}
\no
where
$a^2\equiv a_ia_i$. Because there is no true gauge invariance in \chiral\
(the parameters of the transformation in \chg\ are holomorphic or
antiholomorphic) this is a $D=4$ $\s$-model.
However by changing variables as: $a_1=r \cos \th$, $a_2=r \sin \th$ and after
a few rescalings of the variables and the shifting $\th\to \th +u/2$
one discovers that \cmbp\ reduces to the
$\s$-model obtained in the axial gauging plus the action for a free boson
corresponding to a time-like (space-like) coordinate if $b>0$ ($b<0$).
The spacetime has signature $(+,+,-{\rm sign}(b), -{\rm sign}(b))$.
This is similar to a relationship between axially gauged and chiral
gauged WZW models found in \KSTh\ for the case of simple groups and if
the subgroup is an abelian one.

\newsec{Operator method}

In general there are $\a'\sim 1/k$ corrections to the semiclassical expressions
for the $\s$-model fields one obtains by integrating out the gauge fields,
as we have done so far, in the gauged WZW action.\foot{There exist
regularization schemes in conformal perturbation theory in which the
semiclassical results
for $SL(2,\IR)/\IR$ and $SL(2,\IR) \otimes \IR/\IR$ solve the
$\b$-function equations to two loop order in perturbation theory
\ark\KST\
(for a different argument that such a scheme should exist see \GK).
However, this is far from being a general conclusion
for all gauged WZW models. Nevertheless, the exact expressions,
as ones obtains them by making contact with the exact coset CFT, are needed in
order to correctly describe the Klein-Gordon-type of equations for the
string modes \ark\KST.}
One can find the exact expressions for the metric and dilaton fields
we have obtained in \meax, \rese\ by identifying the underlying exact CFT.
As in was noted in \NW\ using the OPE expansions
\eqn\OPE{\eqalign {
&P_i P_j \sim {\e_{ij}T \ov z-w} +{\d_{ij}\ov (z-w)^2}\ ,\qq
JP_i \sim {\e_{ij}P_j \ov z-w} \cr
&JJ \sim {b\ov (z-w)^2} \ ,\qq JT \sim {1\ov (z-w)^2}\ ,\qq TT\sim 0\ , \cr} }
\no
one can show that the stress energy tensor defined as
\eqn\st{T_G=\ha :( P_i P_i + JT + TJ + (1-b) T^2 ):}
\no
satisfies the Virasoro algebra with central charge $c=4$. In fact this
corresponds to a
solution of the Master equation of \HK\MOR. It also obvious that
$T_H={1\ov 2b} :J^2:$ satisfies the Virasoro algebra with central charge
$c=1$. The difference $T_{G/H}=T_G-T_H$ satisfies the same algebra with $c=3$.
What is also true is that $T_{G/H} J \sim 0$. The regularity of the last OPE
makes it possible to gauge the corresponding symmetry.

It is convenient to express the zero modes of the holomophic currents in \OPE\
(and of the
corresponding anti-holomorphic ones) as first order differential operators.
We compute the `left' and `right' Cartan forms
\eqn\gidg{\eqalign{
&g\inv dg= (\cos u\ da_j - \sin u\ \e_{ij} da_i )\ P_j + du\ J +
(dv + {1\over 2}\e_{ij} da_i a_j)\ T \cr
&dgg\inv= (da_j -\e_{ij} a_i\ du)\ P_j + du\ J
+ (dv-{1\over 2} \e_{ij} da_i a_j -{1\over 2} a_i a_i\ du)\ T \ .\cr } }
\no
{}From the above we compute the following matrices defined as
$g\inv dg = dX^M L_M{}^A T_A$, $-dg g\inv = dX^M R_M{}^A T_A$, where
$X^M=\{a_1,a_2,u,v\}$
\eqn\L{L_M{}^A=\pmatrix{\cos u & -\sin u & 0 & {a_2\over 2} \cr
\sin u & \cos u & 0 & -{a_1\over 2} \cr
0& 0 & 1& 0\cr
0 & 0 & 0& 1\cr} ,\
R_M{}^A=\pmatrix{-1 & 0 & 0& {a_2\over 2}\cr
0 & -1 & 0 & -{a_1\over 2} \cr
-a_2 & a_1& -1 & {1\over 2}(a_1^2 +a_2^2)\cr
0&0&0&-1 } }
\no
and their inverses
\eqn\Li{L_A{}^M= \pmatrix{\cos u & \sin u & 0 &
-{1\over 2}(a_2 \cos u - a_1 \sin u)\cr
-\sin u & \cos u & 0 & {1\over 2}(a_1 \cos u +a_2 \sin u) \cr
0& 0 & 1 & 0 \cr
0 & 0 & 0 & 1\cr } , \
R_A{}^M= \pmatrix{-1& 0 & 0 & -{a_2\over 2}\cr
0 & -1 & 0 & {a_1\over 2} \cr
a_2 & -a_1 & -1 & 0\cr
0 & 0 & 0 & -1 \cr}  }
\no
The first order differential operators defined as
$J_A= L_A{}^M \partial_M$, $\bar J_A= R_A{}^M \partial_M$ satisfy the
commutation relations \alg\ and commute with each other.
Explicitly they are given by
\eqn\lc{\eqalign{&J_{P_i}= (\cos u\ \d_{ij} + \sin u\ \e_{ij})\partial_{a_j}
+{1\over 2} (\sin u\ \d_{ij} - \cos u\ \e_{ij}) a_j \partial_v \ ,\quad
J_J=\partial_u \ ,\quad J_T =\partial_v \cr
& \J_{P_i}= -\partial_{a_i} - \ha \e_{ij} a_j \partial_v\ ,\quad
\J_J= -\e_{ij} a_i \partial_{a_j} -\partial_u \ ,\quad \J_T=-\partial_v\ .
\cr }  }
\no
As usual the metric and the dilaton in any WZW model (gauged or not)
can be deduced by comparing \DVV\BSexa\ $H T= (L_0 +\bar L_0) T $
with
$$HT= -{1\ov \sqrt{G}\ e^{\Phi}}
\partial_{M} \sqrt{G}\ e^{\Phi}\ G^{MN} \partial_{N} T \ ,$$
where $H$ is the Hamiltonian of the corresponding CFT, $L_0$ and $\bar L_0$ are
the zero modes of the holomorphic and antiholomorphic stress energy tensors
and $T$ denotes tachyonic states of the theory annihilated by the positive
modes of the holomorphic and antiholomorphic currents.
For the case of the $D=4$ WZW model the tachyon depends on all af the four
group parameters,
i.e. $T=T(a_1,a_2,u,v)$. In that case the forementioned comparison gives for
the inverse metric
\eqn\mei{G^{MN}=\pmatrix{1 & 0 & 0 &-{a_2\ov 2}\cr
0 & 1 & 0& {a_1\ov 2}\cr
0 & 0 & 0 & 1\cr
 -{a_2\ov 2}&{a_1\ov 2}&1&{a_1^2\ov 4}+{a_2^2\ov4}+1-b\cr} }
\no
which upon inverting it gives the metric corresponding to the $\s$-model
\action, but with $b\to b-1$. One might think of the shifting as a quantum
correction to the semiclassical result in \action. However, this it does not
really matter since one can absorb $b$ into a redefinition of $v$
(however it will be important for the gauged models that we will shortly
consider).
The dilaton turns out to be constant in this case as expected.
For the $D=3$ model case one should choose gauge invariant tachyonic states $T$
because of the gauge symmetry \BSexa.
For the case of the axial gauging we have the
constraint
\eqn\invaa{( J_J- \J_J)T=0 \quad \Rightarrow  \quad
T=T(\r^2=a_ia_i\ ,v\ ,x=a_1\cos {u\ov 2}+a_2 \sin {u\ov 2} )\ .}
Then the result for the inverse metric is (with $X^{\m}=\{v,x,\r \}$)
\eqn\inme{G^{\m\n}=\pmatrix{{\r^2\ov 4}-b+1& 0 & 0\cr
0 & 1+{1\ov 4b}(x^2-\r^2) & {x\ov \r} \cr
0 & {x\ov \r} & 1\cr } \ .}
\no
After we invert it and change variables as $x=\r \cos {u\ov 2}$ we obtain
the metric and dilaton
\eqn\obt{\eqalign{
&ds^2=d\r^2 +{4\ dv^2 \ov \r^2 -4b+4} - {b\ \r^2 \ov \r^2-4b}\ du^2 \cr
&\Phi=\ha \ln\bl( (\r^2-4 b)(\r^2 - 4 b+4) \br) +{\rm const.}\ .} }
The above expressions become equivalent to the ones in \meax\ for large $b$.
To obtain the result of the vector gauging one has to impose the constraint
\eqn\vcon{ (J_J+\J_J)T = 0 \quad \Rightarrow \quad
T=T(\r^2=a_ia_i\ ,u\ , v)\ .}
In this case one obtains
\eqn\tora{ds^2=d\r^2 +{4 dv^2\ov \r^2 +4} - b du^2\ , \qq
\Phi=\ha \ln \bl( \r^2 (\r^2 +4) \br) + {\rm const.}\ .}
These results become equivalent to the ones in \rese\ after we rescale
$\r\to b \r$ and take the large $b$ limit.
For all cases, namely the $D=4$ WZW and the $D=3$ gauged WZW models one can
verify that
indeed the physical condition for closed strings $(L_0-\bar L_0)T=0$ is obeyed.
It can be shown that \obt\ and \tora\ are related to the exact expressions for
the metric and dilaton of the $3D$ charged black string \SFET\foot{
To compute the exact antisymmetric tensor one needs to use an effective
action approach. Assuming that in our case it can also be obtained as a limit
of the corresponding expression in the charged black string case one can
show that both prescriptions of \KST\ give for it the semiclassical expression
of \meax.} and the
$2D$ black hole (times a free boson) \DVV\ through a limiting procedure
similar to the one we described in the previous section.
Therefore their conformal invariance has already been checked against
perturbation theory in \KST\TSEY\JJP.

\newsec{Concluding remarks and discussion}

We considered various gaugings of a one dimensional subgroup of
a WZW model based on a four dimensional non-semi-simple group.
We explicitly demonstrated that the three dimensional $\s$-models we have
obtained can be
derived by taking a correlated limit of models of gauged WZW based on
{\it semi-simple} groups. In particular the limit we took corresponds to
`magnifying' or blowing up the region around the curvature singularities in
the latter models. We show that our backgrounds, although they still have
curvature singularities, can be mapped to flat spacetimes
via duality transformations which renders these singularities harmless.
In contrast the singularities of the original models which were based on
semi-simple groups were more severe (although still not peculiar from the
CFT point of view \WIT\GIN) in the sense that duality transformations
cannot remove them, i.e. both the charged and the neutral black strings have
curvature singularities.

It will be interesting to construct other WZW and gauged WZW models based
on non-semi-simple groups.
If the conclusions we drew by considering the particular examples
of this paper are generalizable certain gauged WZW models based on
non-semi-simple groups describe the geometry of gauged WZW models based
on simple or semi-simple groups close to the curvature singularities the latter
models have. Then by duality transformations we might be able to map these
space (which still have curvature singularities) to non-singular ones.
This in turn is very important in order to understand better
gravitational singularities in the context of String theory.

\bs

\centerline  {{\bf Acknowledgments}}

I would like to thank CERN for providing its computer facilities in the final
stages of typing this paper. I also thank Prof. 't Hooft for a discussion
and C. Nappi for useful remarks.

\centerline{ {\bf Noted added} }

\no
1). While finishing the typing of the paper we received ref.\KK, which contains
some overlapping
materials with the present work. In particular gauging of the same WZW model
was considered, but with a subgroup generated by $P_1$ instead of $J$.
The resulting $\s$-model is different than ours but of course it also has
$c=3$. We were also informed about some relevant work in \KT.

\no
2). After this paper was submitted for publication the paper \ORS\ appeared
where a large class of WZW models based on non-semi-simple groups was
constructed as a particular contraction of the WZW model for $G\otimes H$.
The model of \NW\ corresponds to $G=SO(3)$ and $H=SO(2)$.
That explains why the action \act\ can be obtained from \acts\ through
a limiting procedure. The gauging of the generator $J$ we have been considering
corresponds to a gauging of the total subgroup current in $G\otimes H$ in the
models of \ORS. That again gives an explanation of the relation between,
for instance, \meax\ and \blst. The generalization of the present work to
cover gauged WZW models based on the non-semi-simple models of \ORS\
is currently under investigation.
Also the paper \edc\ appeared where an explicit expression for the WZW action
based on the non-semi-simple group $E^c_d$, i.e. a central extension of the
Euclidean group in $d$-dimensions, was given.

\listrefs
\end